\documentclass[aps,prd,letterpaper]{revtex4}

\usepackage{latexsym}
\usepackage{epsfig}
\usepackage{scrextend}
\usepackage[french]{babel}
\usepackage{url}
\usepackage{hyperref}
\usepackage{latexsym}
\usepackage{epsfig}
\usepackage{scrextend}
\usepackage{lipsum}   

\usepackage{natbib}

\usepackage{tikz}
\usetikzlibrary{positioning,arrows,decorations.pathmorphing,arrows.meta}

\usepackage{xcolor}
\usepackage{amssymb}
\usepackage{amsmath}
\usepackage{graphicx}

 \definecolor{blue}{RGB}{7,80,201}
 \definecolor{red}{RGB}{200,20,1}

\def\be{\begin{equation}}
\def\ee{\end{equation}}
\def\bea{\begin{eqnarray}}
\def\eea{\end{eqnarray}}
\def\ba{\begin{array}} 
\def\ea{\end{array}}
\def\bc{\begin{center}}
\def\ec{\end{center}}

\def\ghost#1{}

\def\simge{\mathrel{
   \rlap{\raise 0.511ex \hbox{$>$}}{\lower 0.511ex \hbox{$\sim$}}}}
\def\simle{\mathrel{
   \rlap{\raise 0.511ex \hbox{$<$}}{\lower 0.511ex \hbox{$\sim$}}}}

\def\mo{\mu_{\circ\hspace{-.2mm}\circ}}
\def\eo{\epsilon_{\circ\hspace{-.2mm}\circ}}

\def\dis{\displaystyle}

\newcommand{\footremember}[2]{%
    \footnote{#2}
    \newcounter{#1}
    \setcounter{#1}{\value{footnote}}%
}

\begin{document}

\title{{\ } 
$\ba{c} \hbox{Completing the International System of units} \vspace{2mm}\\
\hbox{with {\boldmath$\,c=\hbar=\mu_\circ=\epsilon_\circ=k_B=$} {\normalsize\boldmath $N_A$} {\boldmath$=1$}}
\ea $
\vspace{3mm}
 }

\author{{\sc P}{ierre} {\sc Fayet}
\vspace{5mm} \\ \small }

\affiliation{Laboratoire de physique de l'\'Ecole normale sup\'erieure\vspace{.5mm}\\ \hspace*{4mm}24 rue Lhomond, 75231 Paris cedex 05, France 
\hspace{-1mm}\footremember{a}{\vspace*{0mm}   \em LPENS, Ecole normale sup\'erieure, Universit\' e PSL, CNRS,
\vspace{.8mm}Sorbonne Universit\'e,    Universit\' e de Paris
}
\vspace{1.5mm}\\
\hbox{and \,Centre de physique th\'eorique, \'Ecole polytechnique, 91128 Palaiseau cedex, France}\
\vspace{3mm}}

\date{June 10,  2019}  

\begin{abstract}

\textwidth 10cm

\vspace{3mm}

\bc
{\bf Abstract}
\ec

{\vspace{-2.5mm}

\hspace*{2mm}

\vspace{.2mm}
A drawback of the new SI is that by fixing the value of the elementary charge $e$, the vacuum magnetic permeability $\mu_\circ$ and impedance $Z_\circ\!=\mu_\circ c$ are  no longer fixed, but get written proportionately to $\alpha$. 
  All electrical units get dependent on $\alpha$ (and might even, conceivably, vary with time).
 
\vspace{1.3mm}
This may be cured by embedding the SI
in a new framework in which the ``fundamental constants of nature'' are fixed and equal to 1, i.e.~\hbox{$c=\hbar=\mu_\circ=\epsilon_\circ=Z_\circ= k_B=N_A=1$}.
The metre, joule, and kilogram 
\vspace{.1mm}
get {\it identified} as  1\,m =  (1/$c_\circ$) s = (1/299\,792\,458) s, \ 1\,J = $(1/\hbar_\circ)$ $  \rm s^{-1}= (2\pi /6.626\,070\,15) \times 10^{34}\ \rm s^{-1}$ and  
1 kg =  $(c_\circ^2/\hbar_\circ)\ \rm s^{-1}= 0.852\,... \times\, 10^{51}\ \rm s^{-1}$.
\vspace{-.3mm}
Fixing $\mu_\circ\!= \mo \rm\, N/A^2=1 $  provides
1\,A = $\sqrt{\mo \rm N} =\!\sqrt{\mo c_\circ/\hbar_\circ}  \ \rm s^{-1}$ and  
\vspace{-.5mm}
1\,C = $\sqrt{\mo c_\circ/\hbar_\circ} = 1.890\,...\times 10^{18} $, 
with $e = 1.602\,... \times 10^{-19}\ \rm C$ also equal to
$ \sqrt{4\pi\alpha} = 0.3028\,...\,$.
All SI units can be defined in terms of the second, with the coulomb, ohm and weber dimensionless, 
and the mole identified as the very large Avogadro number.

\vspace{0mm} }
\end{abstract}

\maketitle

\vspace{-4mm}

\section{\boldmath Having to cope with an unfixed {\normalsize $\mu_\circ$}\,, \,in the new SI}

The International System of units is based on six fundamental units,  leaving aside the candela \cite{si}.
 \,It has just been  revised so that the kilogram, the ampere, the kelvin and the mole get redefined in terms of fixed numerical values of the Planck constant $h$, the elementary charge $e$, the Boltzmann constant $k_B$, and the Avogadro constant $N_A$ \cite{mills,mills2,rev,rev2,cr}.
All basic units are now defined from reproducible universal phenomena, rather than from unique material artefacts, as were the metre and kilogram stored for years 
at the Bureau International des Poids et Mesures (BIPM) \cite{m,kg}.

\vspace{1.5mm}

But, by deciding to fix the elementary charge to be, exactly, $e = 1.602\,176\;634 \times 10^{-19}$ C, the coulomb, and thus the ampere, have to be slightly adjusted accordingly. 
With $e=e_\circ\,\rm C$  replacing its earlier expression  as  $e=e_{\rm old}\, \rm C_{\rm old}$, one has \cite{pfcr}
\be
\label{ca}
\left\{\,\ba{c}1 \,\rm C= \eta \ C_{\rm old},\
\vspace{2mm}\\
1\, \rm A= \eta \ A_{\rm old}\,, 
\ea\right. \ \
\hbox{and thus }\ \
\left\{\,\ba{ccl}
1 \,\rm V\!&=&  \rm \eta^{-1} \ \ V_{\rm old}\,,
\vspace{1mm}\\
1 \,\rm Wb\!&=& \rm  \eta^{-1} \ Wb_{\rm \,old}\,,
\vspace{1mm}\\
1\, \rm T\!&=& \rm \eta^{-1} \ \ T_{\rm old}\,,   \
\ea\right. \  \ \
\left\{\ \ba{cccc}
1 \,\rm F\!&=&\! \eta^{2}\  &\!\! \rm F_{\rm old}\,,
\vspace{1mm}\\
  1\, \rm H\!&=&\!  \eta^{-2} &\!\!\rm H_{\rm old}\,,
  \vspace{1mm}\\
  1\,\rm\Omega \!&=&\! \eta^{-2}&\!\!\Omega_{\rm old}\,,
\ea\right.
\ee
with $e_\circ = \eta^{-1}\,e_{\circ \,\rm old}$. The numerical value $e_\circ$ of the elementary charge  is chosen very close to its former best determination so that $\eta$ is very close to 1, up to a few $10^{-10}$.

\vspace{1.5mm}
The ampere was previously defined as A$_{\rm old}$, 
that constant current which, if maintained in two straight parallel conductors of infinite length, 1\,m apart in vacuum, would produce between them a force of $\,2 \times 10^{-7}$\;N per metre of length
\cite{amp}. The ampere getting multiplied by $\eta$, the new force per metre of length 
will now be $2\!\times \! 10^{-7}\ \eta^2$ N/m.  If we want to keep the usual expression of Amp\`ere's force law, $F/L= \mu_\circ I^2 /\,2\pi r$, we must abandon the fixed expression for the vacuum magnetic permeability $\mu_\circ$, 
through the replacement
\be
\label{mu}
\mu_\circ= 4\pi\!\times \!10^{-7}\ \rm N/A^2\ \ \ \rightarrow\ \ \  \mu_\circ = \mo\, \rm N/A^2 =4\pi\!\times \!10^{-7}\ \eta^2 \ \rm N/A^2\,.
\ee
The vacuum electric permittivity $\epsilon_\circ= 1/\mu_\circ c^2$ becomes also unfixed,
\be
\label{eps}
\epsilon_\circ = \eo\,{\rm C^2/(J\,m)} = \frac{1}{4\pi\!\times \!10^{-7}\,c_\circ^2}\ \eta^{-2}\ \rm C^2/(J\,m)\,= 8.854\;187\;817\;620\,... \times  10^{-12} \  \eta^{-2} \ F/m\,.
\ee

\vspace{-3mm}

\noindent
$\mo$ and $\eo$ are dimensionless coefficients  related by  $\eo\mo =1/c_\circ ^2\,$,
free to adjust to the new definitions of the ampere and  coulomb.

\vspace{2mm}

As a result {\bf\em \boldmath the sizes of electrical units now depend on $\alpha$}. They are no longer rigidly tied to the mechanical ones. 
The connection is made softer, with a freely-floating $\mo$ as a parameter, so that $\mu_\circ$ may now qualify as a new ``fundamental constant of nature'', on the same ground as $c$ and $\hbar$. 
The fine structure constant gets expressed as 
\be
\label{alpha}
 \alpha= \frac{e^2}{4\pi\epsilon_\circ\hbar c}\,=\,\frac{\mu_\circ c \,e^2}{4\pi\hbar}=
\frac{10^{-7}\,c_\circ \,e_\circ^2}{\hbar_\circ}\ \eta^2
=\alpha_\circ\, \eta^2
= \frac{\eta^2}{137.035\;999\;158\;713\,...}
=\frac{1}{137.035\;999\;084\,(21)}\,.
\ee 
The vacuum magnetic permeability $\mu_\circ$, measured in N/A$^2$ or H/m, is proportional to $\alpha$ and depends on its future measurements, through 
\be
\label{muz}
\mo= \,4\pi\!\times\!10^{-7}\ \eta^2= 4\pi\!\times\!10^{-7}\ \alpha/\alpha_\circ\,=\,12.566\; 370\; 6212\,(19) \times 10^{-7}\,.
\ee
The 2017 adjustment \cite{codata1,codata2}, taken with the exact values of the fixed constants~\cite{ampmise},   
corresponds to
\be
\label{eta}
\eta^2\,= \,1+2.0\,(2.3) \times 10^{-10}\,.
\ee
But with the new 2018 recommended values \cite{alpha} of $\alpha$ and $\mu_\circ$ as expressed in  (\ref{alpha},\ref{muz}) we now obtain a slightly higher value

\vspace{-8mm}
\be
\label{eta1}
\ba{c}
\dis \eta^2\,=\,\frac{\alpha\,}{\alpha_\circ}
\,=\, \frac{\mu_\circ}{4\pi\!\times\! 10^{-7}\ \rm N/A^2} \,=\, 1+    5.4\,(1.5)\times 10^{-10}\,.
\vspace{-1.5mm}\\
\ea
\ee

\vspace{0.5mm}
\noindent
This already shows (at $\approx 3.6\ \sigma$) a small discontinuity in the definition of the new electrical units, as compared to the older ones. It provides back from (\ref{eps}) $\epsilon_\circ=  8.854\;187\;8128\,(13) \times 10^{-12}$ \rm F/m, as in \cite{eps}.

\vspace{1mm}

Similarly the vacuum impedance $Z_\circ=\!\sqrt{\mu_\circ/\epsilon_\circ}=\mu_\circ c$, previously fixed,
now has a slightly higher value
\vspace{-3mm}
\be
\label{z}
\ba{ccl}
Z_\circ
= (4\pi\times10^{-7}\, \eta^2 \rm \ N/A^2) \times (299\,792\,458\  m/s) = 
\mo c_\circ\ \Omega \!&=&\! 376.730\;313\;461\;770\,... \ \eta^2\ \Omega
\vspace{2mm}\\
\!&=&\! 376.730\;313\;665\,(57)\,... \ \Omega\,.
\ea
\ee

\vspace{-1mm}
\noindent
Its value in ohms depends on future measurements of $\alpha$,
even if $\,Z_\circ$ is supposed to be a characteristic of the vacuum.
Fortunately, the ohm itself is now proportional to $\eta^{-2}$ as seen in (\ref{ca}),
so that $Z_\circ= 376.730\,... \,\eta^2\, \Omega\,$ is independent of $\eta$, and $\alpha$  \cite{pfcr}.
This  indetermination of the values of the vacuum impedance (in $\Omega$), magnetic permeability $\mu_\circ$ (in H/m),  and electric permittivity (in F/m) is a new feature introduced in the SI on its last 2019 revision, decoupling electrical units from mechanical ones. 
It may be viewed, conceptually, as an unfortunate feature of the new SI, resulting from the desire to fix the value of $e$ so that $K_J=2e/h$ and $R_K=h/e^2$ be numerically fixed in GHz/V and $\Omega$, respectively. 
\vspace{2mm}

 {\bf\em The electrical units could even depend on time, in the new SI}, 
\,should $\alpha$ depend on time, as occasionally considered.
They would no longer be time-invariant  (in principle a basic requirement for fundamental units)
but could vary, according to
\be
\label{dot}
\rm \frac{\dot C}{C}\,=\, \frac{\dot A}{A}\,=\,  -\,\frac{\dot V}{V}\,=\,  -\,\frac{\dot T}{T}\,=\,-\,\frac{1}{2}\ \frac{\dot\Omega}{\Omega}\,=\,  ... \,=\,   \frac{\dot \eta}{\eta}\,=\,  \frac{1}{2}\ \frac{\dot \alpha}{\alpha}\,
\ee
(with $\rm 1\,C\! \times \!1\,V = 1\,J$ and  $\rm 1\,A \!\times  \!1\,m \!\times \!1\,T = 1\,N\,$ independent of $\eta$ and $\alpha$), even if such variations are constrained to be extremely small (and practically insignificant).
This is the price to pay for fixing $e$ and  leaving $\mu_\circ$ and $Z_\circ=\mu_\circ c$  unfixed. 
Fortunately, as we shall see,  we can still embed the new SI within a larger framework, bringing  in all the advantages of a system  in which $c=\hbar=\mu_\circ=$ $\epsilon_\circ = 1\,$.

\vspace{-2mm}

\section{Fixing  {\normalsize \lowercase{\em \,c}}  = {\normalsize $\hbar$} = {\normalsize $\mu_\circ$} = {\normalsize $\epsilon_\circ$} = {\normalsize 1}\,, \,within the new SI}

\vspace{-1mm}

In contrast with the Syst\`eme International d'unit\'es, revised since May 20th, 2019, theoreticians often like to consider an ``ideal" choice of $c=\hbar=1$, extended to electromagnetism with $\,\mu_\circ\!=\epsilon_\circ\!=1$.
\,But  this is usually considered as a utopic system, as opposed to the generality, practicality and universality 
\linebreak

\pagebreak

\noindent
of  the SI. Still, the shall show how the new SI, as officially redefined, can be included within a larger framework in which $c=1$ as suggested by relativity, and $\hbar =1$ as suggested by quantum mechanics.
$\hbar=1$ is implicitly adopted, in fact, when we say that the electron  is a spin-1/2 particle.

\vspace{2mm}

Then we can again directly view the vacuum magnetic permeability $\mu_\circ$, electric permittivity $\epsilon_\circ$ and impedance $Z_\circ=\sqrt{\mu_\circ/\epsilon_\circ}$ as fixed quantities, as for the speed of light $c=1/\sqrt{\mu_\circ\epsilon_\circ}$ and reduced Planck's constant $\hbar$.  All of them may be regarded  as ``fundamental constants of nature''. 
\vspace{.2mm}
We may then, in addition, fix them to 1, which provides expressions of the metre  in terms of the second (from $c$),
\vspace{-.2mm}
 of the joule in terms of the s$^{-1}$ (from $\hbar$), of the ampere in terms of $\sqrt {\rm N}$ (from $\mu_\circ$), etc..

\vspace{2mm}

The theory of relativity already allowed for defining the metre from the second, by fixing the value of $c$ to be, {exactly},  $299\;792\;458\ \rm m/s\,$ \cite{c}.
If we impose, {\it in addition}, $c=1$ \cite{pfcr}, we can  consider the second as being also a unit of length, and identify the metre as a fixed fraction of the second:
\be
\label{m}
c= c_\circ {\rm \ m/s = 299\;792\;458\ m/s = 1\ \ \ \Longleftrightarrow\ \ \ 1\,m}\ \equiv\ (1/c_\circ) \ \rm s=(1/299\;792\;458)\ s\ .
\ee
The definition of the kilogram has just changed, from the historical ``grand $\cal K$''  stored at BIPM \cite{kg} to a new one based on quantum physics \cite{rev,rev2,cr}. This is done by fixing the value of Planck's constant to
$ h= 6.626\;070\;15 \times 10^{-34}\ \rm J\,s\,.$
If we now impose, {\it in addition,} \,$\hbar=h/2\pi=1$, we can consider the second$^{-1}$ as being, also, a fundamental unit of energy, and identify the joule as a fixed number of \,s$^{-1}$:
\be
\label{J}
\hbar= \hbar_\circ\, \rm J\,s=(6.626\;070\;15 \times 10^{-34}/2\pi)\ \rm J\,s=1\ \  \Longleftrightarrow\ \ 
1\,J \equiv (1/\hbar_\circ)\ s^{-1}= 0.948\;252\;156\;246\,... \times 10^{34}\  \hbox{s}^{-1}.
\ee

\vspace{-1mm}

\noindent
The newton and the kilogram can then be identified as
\be
\label{Nkg}
\left\{\ 
\ba{ccccccc}
{\rm 1\,N} \!&=&\!  \rm   1\ J/m \!&=&\! {c_\circ }/{\hbar_\circ}\ \rm s^{-2}\!&=& \! 2.842\;788\;447\;250\,... \times 10^{42}\ \, \hbox{s}^{-2} \,, \vspace{2mm}\\
{\rm 1\,kg}\!&=&\! 1\ \rm J\,s^2\!/m^2\!&=&\!{c_\circ^2 }/{\hbar_\circ}\ \rm s^{-2}\!&=&\! 0.852\;246\;536\;175\,... \times 10^{51}\ \, \hbox{s}^{-1} \,.
\ea \right.
\ee

\vspace{1mm}

But can we go further in our plan to include the new SI within a system where $c=\hbar=\mu_\circ=\epsilon_\circ=1$ ? 
This may seem difficult, especially as $\mu_\circ$ is no longer exactly fixed but given by $\mu_\circ = 4\pi\times 10^{-7}\,\eta^2\ \rm N/A^2$, its value being now dependent on $\alpha$.   
Irrespectively of this, we can impose $\mu_\circ=1$, which provides the ampere as proportional to a square root of the newton:
\be
\label{mu1}
\mu_\circ=\mo\  \rm N/A^2= 4\pi\!\times \!10^{-7}\ \eta^2\ \rm N/A^2\,=1\ \ \ \Longleftrightarrow\ \ \ 1\,A=\sqrt {\mo\,\rm N}=\sqrt {4\pi \!\times \!10^{-7}\,\rm N}\ \,\eta\,.\ \ \ 
\ee 
With $c=1$, asking for $\mu_\circ=1$ or  $\epsilon_\circ = 1/\mu_\circ c^2=1$ is equivalent, with
\be
\label{eps1}
\hspace{.15mm}
\epsilon_\circ =\eo{\rm \,C^2/(J\,m)}
= \frac{1}{4\pi\!\times \!10^{-7}\,c_\circ^2}\ \eta^{-2}\ {\rm C^2/(J\,m) =1\  \ \Longleftrightarrow\ \
1\,C}=\sqrt{{\rm J\,m}/\eo}=\sqrt {4\pi\!\times \! 10^{-7}\,c_\circ^2\,\rm J\,m}\ \eta\,,\!\!
\ee
verifying from (\ref{mu1},\ref{eps1}) that \,1\,C = $ c_\circ$\,m\,A = 1\,A\,s.
This is also equivalent to choosing in (\ref{z})
\be
\label{z1}
Z_\circ=\sqrt{{\mu_\circ}/{\epsilon_\circ}} =\,\mu_\circ c \,=\, 4\pi\!\times \!10^{-7}\,c_\circ \ \eta^2\ \Omega\,=
\,376.730\,313\,461\,770\,...\  \eta^2\ \Omega=\,1\,.
\ee

The new SI gets embedded within a larger framework, so that 
$\mu_\circ,\ \epsilon_\circ$ and $Z_\circ =\sqrt {\mu_\circ/\epsilon_\circ}=\mu_\circ c$ \,are all fixed and equal to 1\,.
The ohm is a dimensionless unit,

\vspace{-4mm}

\be
1\,\Omega\,={1}/{\mo c_\circ}= {1}/{(376.730\,...\!\times \eta^2)}\ ,
\ee
as required for
$Z_\circ=\mo\,c_\circ\ \Omega\,$ in (\ref{z},\ref{z1}) to be independent of $\alpha$, and equal to 1, the natural unit of impedance. The ampere and the coulomb are obtained from (\ref{mu1},\ref{eps1}) as
\be
\left\{ \,\ba{cccclcl}
\rm 1\,A\!&=&\! \rm \sqrt{\mo\,N}\!&=&\!  \sqrt{\mo c_\circ/\hbar_\circ}\ \rm s^{-1}\!&=&\!  1.890\;067\;014\;853\,...\times 10^{18}  \ \eta\  \rm s^{-1} , 
\vspace{2mm}\\ 
\rm 1\,C\!&=&\! \rm1\,A\,s\!&=&\!  \sqrt{\mo c_\circ/\hbar_\circ} \!&=&\!  1.890\;067\;014\;853\,...\times 10^{18} \ \eta 
\ =\,1.890\;067\;015\;36\,(14)\,...\times 10^{18}.
\ea \right.
\ee
With $e=e_\circ \, \rm C= 1.602\;176\;634 \times 10^{-19}\, \rm C$ and $\alpha= e^2/4\pi$ we recover
\be
e = \sqrt{\mo c_\circ\,e_\circ^2/\hbar_\circ} = \sqrt{4\pi\alpha}=  \sqrt{4\pi\alpha_\circ}\ \eta\ =\ 0.302\;822\;120\;789\,...
\ \eta= 
\,0.302\;822\;120\;871\,(23)\,.
\ee

The weber is also dimensionless, with
\be
{\rm 1\ Wb= 1\,C\times 1\,\Omega }= 1/\sqrt{\mo c_\circ\hbar_\circ}= \,5.017\;029\;284\;119\,...  \times 10^{15}\ \eta^{-1}
= \,5.017\;029\;282 \;76\, (38)\,...  \times 10^{15},
\ee

\pagebreak

\noindent
reflecting that 1\,V = 1\,A\,$\times \,1\,\Omega$.
The flux quantum, $h/2e=\pi/e$, has a fixed value when expressed in Wb, 
\be
\Phi_\circ= \frac{h}{2e}=\frac{\pi \hbar_\circ}{e_\circ} \ \rm Wb = 
2.067\;833\;848\;461\,... \times 10^{-15}\ Wb = \frac{\pi}{e} =\sqrt{\frac{\pi}{4\alpha}}\,\simeq  10.374\,382\,972\,... \ \eta^{-1}\,,
\ee
but is proportional to $\eta^{-1}$ i.e. to $1/\sqrt \alpha$, as for the weber.
The Josephson and von Klitzing constants, numerically fixed in Wb and GHz/V in the new SI \cite{ampmise}, are

\vspace{-4mm}

\be
\!\left\{\!\ba{cccccccccccc}
\dis K_J =\frac{1}{\Phi_\circ}\!
 &=&\! \dis \frac{e_\circ}{\pi\hbar_\circ} &\!\rm Hz/V \!\!
 &=& \! \! 483\;597.\,848\;416\;983\,...& \! {\rm GHz/V}
\!\!&=&\! \dis
\frac{e }{\pi}  =\sqrt{\frac{4\alpha}{\pi}}\!
&=& \!.0963\;912\;748\;023\;449\,...\ \eta\,,
\\ [4mm]
\dis R_K =\, \frac{h}{e^2}\!\!&=& \, \dis \frac{h_\circ}{e_\circ^2}&\! \!\!\Omega\  &=&25\,812.\,807\;459\;304\,... &\Omega &=& \dis
\!\dis \!\!\frac{1}{2\alpha}\!&=& 68.517\;999\;579\;356\,...\ \eta^{-2} .
\ea\right.
\ee
\vspace{-3.5mm}

\noindent
To illustrate better how $R_K$ involves  the vacuum impedance $Z_\circ\!=\mu_\circ c=1$ and $\alpha=\eta^2\,\alpha_\circ$, 
we can write 
\be
\ba{ccl}
\dis R_K= \frac{h}{e^2} 
\!&=&\!
\dis \mu_\circ c\ \,\frac{1}{2}\ \,\frac{4\pi\epsilon_\circ\hbar c}{e^2}
= \,\frac{Z_\circ}{2\alpha} \,= \,\frac{1}{2}\ [Z_\circ \!= 376,730\,...\, \eta^2 \, \Omega=1]\ \eta^{-2}
\left[\,\frac{1}{ \alpha_\circ}= 137.035\,999\,...\,\right] = \frac{1}{2\alpha} 
\vspace{2mm}\\
&=& 25\,812.\,807\,459\,304\,...\ \Omega\ .
\vspace{-4mm}\\
\ea
\ee

\vspace{2.5mm}

The remaining electrical units may be expressed in terms of the second as \cite{pfcr}
\be
\left\{\ 
\ba{ccccccrc}
\rm 1\,V&=&\rm  1\,J/C = 1 \ Wb\ s^{-1}\ \ \ \ &=&   \dis
{1}/{\sqrt{\mo c_\circ \hbar_\circ}}\  \,{\rm s}^{-1}&=&  5.017\;029\;284\;119\,...  \times 10^{15}&\! \eta^{-1}\ \,{\rm s}^{-1}\,,
\vspace{2mm}\\
1\,\hbox{V/m}&=&
\,  1\,{\rm N/C} \  
\!&=&\! \dis \sqrt{{c_\circ}/{\mo\hbar_\circ}}\ \ {\rm s}^{-2}
\!&=&\dis  1.504\;067\;540\;944\,... \times  10^{24}&\!\eta^{-1}\ {\rm s}^{-2}\,,
\vspace{2mm}\\
 1\,\hbox{T}&=&
1\,{\rm N/(A\,m)}\!&=&\! \sqrt{c_\circ^3/\mo\hbar_\circ}\ \ {\rm s}^{-2}
\!&=&\dis  4.509\;081\;050\;976\,... \times  10^{32}&\! \eta^{-1}\ {\rm s}^{-2}\,,
\\ [2mm]
1 \ \hbox{F}&=&\dis 1\,{\rm C/V\,=\,1\,s/\Omega}
&=&\mo c_\circ\ \ {\rm s}
\!&=&\! \ \  376.730\;313\;461\;770\,...\ & \!\!\!\!\!\eta^{2}\ \ {\rm s}\,,\ 
\\ [2mm]
1\  \hbox{H}&=&
1\ \hbox{J/A}^2=\,1\ \Omega\,{\rm s}&=&\dis ({1}/{\mo c_\circ})\ \ {\rm s}
\!&=&\!\dis\!\! 1/376.730\;313\;461\;770\,...\ &\! \!\!\!\!\eta^{-2}\ \, {\rm s}\,.\ 
\ea \right.
\ee
$\eta=1+2.7\,(.8)\times 10^{-10}$, very close to 1 as seen in (\ref{eta},\ref{eta1}), originates from fixing $e$ in view of fixing $K_J$ and $R_K$,  resulting in electric units no longer rigidly tied to the mechanical ones.
This could have been avoided by conserving the earlier 
definitions of the electrical units, with $\eta=1$ and 
$  \mu_\circ = 4 \pi \! \times \! 10^{-7} \ \rm N/A^2$.
$K_J$ and $R_K$ would then have been left unfixed. 
Still they are related, independently of $e$, by
\be
K_J^2\ R_K\,=\,4/h\,=\,2/\pi\,=(4/h_\circ)\ {\rm (\Omega/V^2 s^2)} = 4/(h_\circ \,\rm J\,s)\,=\,.603\;676\;071\;856\;860\,... \times 10^{34}\ (\rm J\,s)^{-1} \,.
\ee

\vspace{.5mm}

Finally we may also fix $k_B=k_{B\circ}\  \rm J/K=1$, $N_A=N_{A\circ}\ \rm mol^{-1}=1$, which define the kelvin and the mole as 
\be
\left\{\ 
\ba{ccccl}
{1\,\rm K }\!&=&\! k_{B\circ} \,{\rm J }= (k_{B\circ}/\hbar_\circ)\ {\rm s}^{-1}\!&=&1.\,380\;649 \times 10^{-23} \ \rm J
= 1.309\;203\;391\;269\,...Ê\times 10^{11}\ s^{-1}\,,
\vspace{2mm}\\
{\rm 1\,mol} \!&=&\! \ \ \ \ \ \ N_{A\circ} \!&=& 6.022\;140\;76 \times 10^{23}\,.
\ea \right.
\ee
The resulting expression of the Kelvin is in agreement with
\be
\ba{ccl}
\dis 1 \,e {\rm V} = e_\circ\,{\rm J} = \frac{e_\circ}{\hbar_\circ}\ \rm s^{-1}\!&=&\! \rm 1.602\,176\,634 \times 10^{-19}\  J = 1.519\,267\,447\,878\,...\,\times 10^{15} \  s^{-1}
\vspace{0mm}\\
&=&\!(e_\circ/k_{B\circ})\ \rm K 
=11\,604.518\,121\,550\,...\ K\,,
\ea
\ee
and the mole gets identified with the very large Avogadro number, $N_{A\circ}$.

\vspace{-1mm}

\section{conclusion}

\vspace{-1mm}

The new SI brought a huge progress by providing a new definition of the kilogram based on quantum physics.
Still by fixing the value of $e$, it requires to adjust the coulomb and ampere, changing the vacuum magnetic permeability $\mu_\circ$ and impedance $Z_\circ=\mu_\circ c$  into unfixed parameters, to be measured experi\-mentally. All electrical units become dependent on $\alpha$, and might even vary (very slightly) with time.

\vspace{1mm}

Still the new SI can be embedded within a larger framework, with $\hbar=c=\mu_\circ=\epsilon_\circ=k_B=N_A=1$, the vacuum magnetic permeability, electric permittivity and impedance being fixed and equal to 1. All units, metre, joule, kilogram, ampere, ... and kelvin get fixed and defined in terms of the second, with the ohm, coulomb and weber dimensionless, $e= 1.602\,... \times 10^{-19}\ \rm C= \sqrt{4\pi\alpha}= 0.3028\,...$, and the mole equal to the Avogadro number.

\pagebreak

\end{document}